# Quantum Theory of Light in a Dispersive Structured Linear Dielectric: a Macroscopic Hamiltonian Tutorial Treatment

Michael G. Raymer
Nov. 26, 2019; updated Jan. 7 2020
to appear in J of Modern  Optics

Oregon Center for Optical, Molecular & Quantum Science, and Department of Physics,

University of Oregon, Eugene, OR 97403, USA

raymer@uoregon.edu

**Abstract**

These notes—intended to be self-contained and tutorial—present a direct, macroscopic approach to quantizing light inside a linear-response dielectric material when both spectral dispersion and spatial nonuniformity are present, but the spectral region of interest is optically transparent so that explicit treatment of the underlying physics of the medium is not needed. The approach taken is based on the macroscopic Maxwell equations and a corresponding Hamiltonian, without the use of Lagrangians or any dynamical model for the medium, and uses a standard mode-based quantization method. The treatment covers: energy density and flux in a dispersive dielectric; a summary of the inverse permittivity formalism; a new derivation of the mode normalization condition; a direct proof of the nonorthogonality of the modes; examples of quantized field expressions for the general case and various special cases; the relationship between group velocity and energy flux; the band approximation and the continuum limit; and quantum optical treatment of waveguide modes.

## 1. Introduction

Strictly speaking, there can be no 'photons' inside a dense dielectric material. Photons are excitations of the EM field in vacuum, and involve only EM energy. Instead, inside a dielectric there are collective excitations of the EM and matter fields together, called polaritons. The energy they store and transport involves both EM and material energy, which are inherently coupled. If an atom embedded inside a dielectric emits a polariton, which is subsequently absorbed by a detector embedded in the same medium, no 'photon' is ever involved in the process. We will, of course, continue to use loosely the term 'light' for what propagates freely inside a dielectric.

The present treatment is meant to be tutorial. It explores a straightforward means for quantizing such a system when both spectral dispersion and spatial nonuniformity are present, but the spectral region of interest is optically transparent so explicit treatment of the underlying dynamics of the medium is not needed. The approach taken is based on the macroscopic Maxwell equations and a corresponding Hamiltonian, without the use of Lagrangians, in an effort to have a mathematically simple derivation that parallels the standard mode-based quantization method presented in most introductory textbooks. The treatment is limited to linear-response materials. New results include a direct derivation of the mode normalization condition, and a simple proof of the nonorthogonality of the modes.

There have been many formulations of the quantum theory of light inside a dielectric. Especially relevant summaries of various approaches are given in (1, 2), and the theory presented here derives from those and other sources, with some original contributions. In dielectrics there are various scenarios, which bring different levels of complexity to the theory, as reviewed in (1, 2, 3, 4). A transparent medium may be spatially inhomogeneous but nondispersive (that is, the refractive index is not frequency dependent) over the spectral range of interest. Or a medium may be spatially homogeneous but dispersive (that is, the refractive index is frequency dependent) over the spectral range of interest. Third, a medium may be both spatially inhomogeneous and dispersive over the spectral range of interest, the case treated here. Finally, a medium may also support significant absorption of light in the spectral range of interest, a regime not treated here.

A careful treatment of field quantization in a linear-response, spatially inhomogeneous but nonabsorbing and nondispersive medium was given by Glauber and Lewenstein (5), who harked back to studies by Born and Infeld (6). One of the earliest careful treatments of quantization in a spatially inhomogeneous and dispersive medium was by Drummond (7), who generalized the method developed by Hillery and Mlodinow (8) for treatment of nonlinear optical media. The mentioned approaches use Lagrangians to ensure correct identification of the canonically conjugate dynamical variables. They use a dynamical model in which the medium response is represented by a collection of quantized harmonic oscillators. Also considering the linear case, Sipe et al (9), and Bhat and Sipe (10) developed a dynamical model that uses only Hamiltonians, not Lagrangians, to model correctly a transparent, homogeneous linear-response dielectric using a dynamical model for the medium. Milonni (11) showed that in the transparent case, the same results can be obtained using an approach in which only the macroscopic Maxwell fields are treated as dynamical, with no dynamical model for the medium.

Here, Milonni's Lagrangian-free, macroscopic approach for a homogeneous dispersive medium is generalized to include spatial structure, in combination with aspects of the Drummond/Hillery/Sipe approaches regarding mode normalization. The approach here is macroscopic, and only Hamiltonians, not Lagrangians, are used. The correctness of the final results is ensured by comparing to the aforementioned treatments.

Because this article focuses on media that are transparent in the spectral range of interest, the complications introduced by losses and absorption will be neglected. It is well known that losses are always accompanied by fluctuations, via a fluctuation-dissipation theorem. Such cases with loss and fluctuations have been treated and reviewed in detail in (3) and many references therein, including the seminal study (12), but are beyond the scope of the present treatment. The method of Bhat and Sipe (10), mentioned above, as well as that of Juzeliūnas (13), also models absorptive dielectric media in the non-transparent frequency range. See also (14) for a recent approach.

Two points are of special interest here: 1) The resulting expressions for the quantized fields, given as weighted sums of bosonic operators, must be consistent with the fundamental theorem for EM energy flow in the medium (Poynting's theorem); and 2) The resulting quantum formalism should be ready for the inclusion of nonlinear-optical effects. With respect to the first point, it is worth noting that some expressions commonly used in the nonlinear optics literature are not consistent with Poynting's theorem, although phenomenological quantization procedures have been used to derive quantum fields that necessarily satisfy Poynting's theorem. In this regard, Blow, Loudon et al (15) is an excellent touchstone, and (16) offers an approximate approach suitable for quantum optics in waveguides.

Regarding the second point, Hillery and Mlodinow (8) and Drummond and Hillery (17) showed that the appropriate fields to treat as 'fundamental' (in the sense of *effective field theories*) in a quantum mechanical nonlinear-optical theory are $D$ and $B$, not $E$ and $B$. The arguments are reviewed clearly in Drummond and Hillery (2) and in Quesada and Sipe (18), which has the provocative title, "Why you should not use the electric field to quantize in nonlinear optics." For this reason, and others going back to Born and Infeld (6), it is more appropriate to use the displacement field $D$ as the 'fundamental' field (that is, a field written as a weighted linear sum of bosonic operators) even for the case of linear optics in a medium. It's worth noting that the more common approach is to regard $E$ and $B$ as fundamental in the sense that they are the fields that determine forces on charged particles in a microscopic theory. But in the macroscopic theory, as treated here, and commonly used in nonlinear optics, $D$ and $B$ play special roles and, thus will be employed here and considered as 'fundamental.'

There are cultural barriers in the nonlinear-optics community to adopting the Born and Infeld quantization approach using $D$ and $B$. The vast majority of papers and books published since Bloembergen's and collaborators' development of nonlinear optics theory (19) have used $E$, not $D$, as a 'fundamental' field. Nearly all papers and books use the forward relation between electric and displacement fields, $D = \varepsilon E$, not the inverse $E = \eta D$, and great efforts have gone into deriving and measuring the electric permittivity $\varepsilon(\mathbf{r}, \omega)$. While it is straightforward to convert between the two forms (2), it does require extra labor, and so practitioners ask, "Why should I?" While there are detectable quantitative differences in the predictions of the two theories in the nonlinear case, they don't show up in strong ways in most experimental situations. But there are cases in which wholly wrong qualitative errors do obtain from naively quantizing $E$ rather than $D$, for example, the sign of the third-order nonlinear-optical frequency shift in a cavity is predicted incorrectly (20).

This note derives from a quantum optics course the author taught for some years at the University of Oregon. The treatment is meant to be tutorial and accessible, using familiar mode expansions and a minimum level of advanced mathematics, such as canonical field variables, field action or Lagrangians. It begins discussing the energy density in a dispersive dielectric. It then summarizes the inverse permittivity formulation of linear optics. Then a theory of mode expansions is given, with emphasis on mode normalization and the *lack* of mode orthogonality in a dispersive, structured dielectric, giving a new, macroscopic derivation of these results. Next, the quantization of the polariton field is treated in the macroscopic formulation, using the derived modes. Some special limiting cases are then summarized, followed by discussion of the continuum limit. Finally, the results are applied to quantization of the field in dielectric optical waveguides, important for emerging quantum technologies.

## 2. The Maxwell equations

The Maxwell equations in a general dielectric medium are:

$$\partial_t \mathbf{B} = -\nabla \times \mathbf{E} \quad ; \quad \partial_t \mathbf{D} = \nabla \times \mathbf{H}$$
$$\nabla \cdot \mathbf{B} = 0 \qquad ; \quad \nabla \cdot \mathbf{D} = 0 \tag{1}$$

ensuring transversality of $B$ and $D$ (whereas $E$ generally is not transverse). As in Eq.(1), we usually omit the space-time arguments $(\mathbf{r}, t)$ if they are clear by context. We use the Fourier transform convention:

$$\mathbf{F}(\mathbf{r}, t) = \int_{-\infty}^{\infty} d\!\!\!/\omega \, e^{-i\omega t} \, \boldsymbol{\mathscr{F}}(\mathbf{r}, \omega) \quad , \quad \boldsymbol{\mathscr{F}}(\mathbf{r}, \omega) = \int_{-\infty}^{\infty} dt \, e^{i\omega t} \, \mathbf{F}(\mathbf{r}, t) \tag{2}$$

where $d\!\!\!/\omega$ stands for $d\omega / 2\pi$, and where the symmetry $\boldsymbol{\mathscr{F}}(\mathbf{r}, -\omega) = \boldsymbol{\mathscr{F}}^*(\mathbf{r}, \omega)$ is needed to ensure $F(\mathbf{r}, t)$ is real. Frequency-domain functions are usually denoted using script fonts. The constitutive relations for a linear-response dielectric are most simply stated in the frequency domain:

$$\boldsymbol{\mathscr{D}}(\mathbf{r}, \omega) = \varepsilon(\mathbf{r}, \omega) \cdot \boldsymbol{\mathscr{E}}(\mathbf{r}, \omega) \; ; \; \boldsymbol{\mathscr{B}}(\mathbf{r}, \omega) = \mu(\mathbf{r}, \omega) \cdot \boldsymbol{\mathscr{H}}(\mathbf{r}, \omega) = \mu_0 \boldsymbol{\mathscr{H}}(\mathbf{r}, \omega) \tag{3}$$

where for an anisotropic material the electric permittivity $\varepsilon(\mathbf{r}, \omega)$ is understood generally to be a tensor quantity, while here the magnetic permeability is assumed a constant scalar, $\mu_0$. Throughout most of this note both will be considered as scalars for simplicity of exposition. It is significant that, because of dispersion (the frequency dependence of the electric permittivity), after transforming Eq.(3) back into the time domain, the functional relation between $E$ and $D$ is not local in time.

### 3. Energy density and flux in a dispersive dielectric

For quantizing the field in a dielectric using a macroscopic Hamiltonian approach, expressions for the magnetic and electric energy of the system are needed. A not-so-widely-discussed fact is that the expression for energy density in a dispersive medium is different than in a nondispersive one. The simple reason is that Poynting's theorem for energy flux in such a medium reflects the fact that energy flows at the group velocity, not the phase velocity. A volume of energy flowing with a given speed creates a flux proportional to (energy density) × (speed). Often the concept of group velocity is associated with finite pulses or wave packets, but even in the limit of a monochromatic field, energy flows at the group velocity, as we will show (and is well known).

Poynting's theorem for energy flux arises in the standard way, valid for an arbitrary linear-response dielectric medium. Define the Poynting vector as $\mathbf{S} = \mathbf{E} \times \mathbf{H}$. Using a standard vector identity, along with the Maxwell equations, gives:

$$\begin{aligned} \nabla \cdot \mathbf{S} &= \mathbf{H} \cdot (\nabla \times \mathbf{E}) - \mathbf{E} \cdot (\nabla \times \mathbf{H}) \\ &= -\mathbf{H} \cdot \partial_t \mathbf{B} - \mathbf{E} \cdot \partial_t \mathbf{D} \\ &\equiv -\partial_t W \end{aligned} \tag{4}$$

which implies the energy density is:

$$W = \int_{-\infty}^{t} dt' \left( \mu_0^{-1} \mathbf{B} \cdot \partial_{t'} \mathbf{B} + \mathbf{E} \cdot \partial_{t'} \mathbf{D} \right) \doteq W_B + W_E \tag{5}$$

where the magnetic energy density is identified as

$$\begin{aligned} W_B &= \mu_0^{-1} \int_{-\infty}^{t} dt' \, \mathbf{B} \cdot \frac{\partial \mathbf{B}}{\partial t'} = \mu_0^{-1} \int_{0}^{B} \mathbf{B} \cdot \partial \mathbf{B} = \frac{1}{2\mu_0} \mathbf{B}^2 \\ &= \frac{1}{2\mu_0} \int d\omega \int d\omega' \, e^{-i(\omega-\omega')t} \boldsymbol{\mathcal{E}}^*(\mathbf{r}, \omega') \cdot \boldsymbol{\mathcal{E}}(\mathbf{r}, \omega) \end{aligned} \tag{6}$$

Here we used the concept of an exact differential to carry out the integration. In contrast, because of dispersion, we cannot carry out an analogous exact integration for the electric energy density, which is local in space, but is not a local-in-time function of either $\mathbf{E}$ or $\mathbf{D}$ alone. Instead, we use Fourier transforms, and find for the electric energy density (for frequencies at which absorption and loss are negligible):

$$\begin{aligned} W_E &= \int_{-\infty}^{t} dt' \, \mathbf{E} \cdot \partial_{t'} \mathbf{D} \\ &= \frac{-i}{2} \int_{-\infty}^{t} dt' \int d\omega \int d\omega' \, e^{-i(\omega-\omega')t'} \left[ \omega \varepsilon(\omega) - \omega' \varepsilon(\omega') \right] \boldsymbol{\mathcal{E}}^*(\mathbf{r}, \omega') \cdot \boldsymbol{\mathcal{E}}(\mathbf{r}, \omega) \\ &= \frac{1}{2} \int d\omega \int d\omega' \, e^{-i(\omega-\omega')t} \left[ \frac{\omega \varepsilon(\omega) - \omega' \varepsilon(\omega')}{\omega - \omega'} \right] \boldsymbol{\mathcal{E}}^*(\mathbf{r}, \omega') \cdot \boldsymbol{\mathcal{E}}(\mathbf{r}, \omega) \end{aligned} \tag{7}$$

To obtain the second line, we write the integral as the sum of two identical, equal terms, and in both we use $\mathscr{E}(\mathbf{r}, -\omega) = \mathscr{E}^*(\mathbf{r}, \omega)$ and the fact that $\varepsilon(-\omega) = \varepsilon(\omega)$ is real in a nonabsorptive (transparent) region of the spectrum, to replace $\omega \to -\omega$, or $\omega' \to -\omega'$, as in (21). To obtain the third line we inserted a regularizing factor $\exp(-\delta t')$, integrated, then took the limit $\delta \to 0$. Rosa, Dalvit and Milonni (21) reviewed the above derivation, attributed to Barash and Ginzberg (22).

To illustrate how the group velocity enters the energy-density expression, consider the common case of a plane wave of incoherent, stationary (steady-state) light traveling in the $z$-direction in a *nonstructured* medium. The medium's refractive index is $n(\omega) = \sqrt{\varepsilon(\omega)/\varepsilon_0}$. The wave's propagation constant is:

$$\begin{aligned} k &= \omega n(\omega)/c \\ &= (\omega/c)\sqrt{\varepsilon(\omega)/\varepsilon_0} \end{aligned} \tag{8}$$

For a stationary random process, all frequency components are statistically independent of all others (as easily verified by Fourier transforms). Therefore, assuming the light is $x$-polarized, we can write:

$$\begin{aligned} \mathscr{E}(\underline{r}, \omega) &= \mathbf{x}\, \mathscr{A}(\omega) e^{ikz} \\ \left\langle \mathscr{A}^*(\omega) \mathscr{A}(\omega') \right\rangle &= I_A(\omega) 2\pi \delta(\omega - \omega') \\ \left\langle \mathscr{E}^*(\underline{r}, \omega) \cdot \mathscr{E}(\underline{r}, \omega') \right\rangle &= I_A(\omega) 2\pi \delta(\omega - \omega') \end{aligned} \tag{9}$$

where $\mathbf{x}$ is a unit polarization vector, $\left\langle \ldots \right\rangle$ is a statistical ensemble average and $I_A(\omega)$ is the spectral density (intensity) of the light. Likewise, from the Maxwell equation in the frequency domain $\left[ i\omega \mathscr{B}(\mathbf{r}, \omega) = \nabla \times \mathscr{E}(\mathbf{r}, \omega) \right]$, the magnetic field is:

$$\mathscr{B}(\mathbf{r}, \omega) = \mathbf{y}\frac{1}{c}\sqrt{\frac{\varepsilon(\omega)}{\varepsilon_0}}\, \mathscr{A}(\omega) e^{ikz} \tag{10}$$

Thus,

$$\left\langle \mathscr{B}^*(\mathbf{r}, \omega) \cdot \mathscr{B}(\mathbf{r}, \omega') \right\rangle = \frac{\varepsilon(\omega)}{c^2 \varepsilon_0} I_A(\omega) 2\pi \delta(\omega - \omega') \tag{11}$$

The ensemble averages of the electric and magnetic energy densities are then:

$$\langle W_E \rangle = \frac{1}{2} \int d\omega \int d\omega' e^{-i(\omega-\omega')t} \left[ \frac{\omega \varepsilon(\omega) - \omega' \varepsilon(\omega')}{\omega - \omega'} \right] \langle \boldsymbol{\mathcal{E}}^*(\mathbf{r},\omega') \cdot \boldsymbol{\mathcal{E}}(\mathbf{r},\omega) \rangle$$

$$= \frac{1}{2} \int d\omega \, I_A(\omega) \left\{ \lim_{\omega' \to \omega} \frac{\omega \varepsilon(\omega) - \omega' \varepsilon(\omega')}{\omega - \omega'} \right\}$$

$$= \frac{1}{2} \int d\omega \, I_A(\omega) \left\{ \frac{\partial(\omega \varepsilon)}{\partial \omega} \right\} \tag{12}$$

$$\langle W_B \rangle = \frac{1}{2} \int d\omega \frac{1}{\mu_0} \langle \boldsymbol{\mathcal{B}}^*(\mathbf{r},\omega) \cdot \boldsymbol{\mathcal{B}}(\mathbf{r},\omega') \rangle$$

$$= \frac{1}{2} \int d\omega \, I_A(\omega) \frac{\varepsilon(\omega)}{c^2 \mu_0 \varepsilon_0}$$

Combining the electric and magnetic parts, and using $c^2 \mu_0 \varepsilon_0 = 1$, the ensemble average of the total energy density is:

$$\langle W \rangle = \frac{1}{2} \int d\omega \, I_A(\omega) \left( \varepsilon(\omega) + \frac{\partial(\omega \varepsilon)}{\partial \omega} \right)$$

$$= \int d\omega \, I_A(\omega) \left( \varepsilon(\omega) + \frac{\omega}{2} \frac{\partial \varepsilon}{\partial \omega} \right) \tag{13}$$

Rosa *et al* pointed out that in this case of uncorrelated frequencies (or for monochromatic light) this result is *exact*, (21) whereas conventional derivations often use a Taylor-series expansion to obtain it, leaving ambiguous the question of exactness (19, 23, 24). To relate this formula to the group velocity $\mathrm{v}_g$ and phase velocity $\mathrm{v}_p$, note that

$$\frac{1}{\mathrm{v}_g} = \frac{\partial k}{\partial \omega} = \frac{1}{c} \left( n(\omega) + \omega \frac{\partial n}{\partial \omega} \right) = \frac{1}{c\sqrt{\varepsilon_0 \varepsilon(\omega)}} \left( \varepsilon(\omega) + \frac{\omega}{2} \frac{\partial \varepsilon}{\partial \omega} \right)$$

$$\mathrm{v}_p = \frac{c}{n(\omega)} = c\sqrt{\frac{\varepsilon_0}{\varepsilon(\omega)}} \tag{14}$$

The ratio of these two velocities is:

$$\frac{\mathrm{v}_p}{\mathrm{v}_g} = \frac{1}{\varepsilon(\omega)} \left( \varepsilon(\omega) + \frac{\omega}{2} \frac{\partial \varepsilon}{\partial \omega} \right)$$

$$= \left( 1 + \frac{\omega}{n(\omega)} \frac{\partial n(\omega)}{\partial \omega} \right) = \frac{1}{n(\omega)} \frac{\partial \omega n(\omega)}{\partial \omega} \tag{15}$$

Therefore, in this case the total energy density can be written:

$$\langle W \rangle = \int \!\!d\omega \, I_A(\omega) \varepsilon(\omega) \frac{\mathrm{v}_p(\omega)}{\mathrm{v}_g(\omega)} \tag{16}$$

Recognizing that the spectrum $I_A(\omega)$ is proportional to electric-field-squared, and that for a plane wave in a dispersionless medium the magnetic energy is equal to the electric energy (as can be seen from results above), we see that Eq.(16) is analogous to the more common expression for energy density,

$$W = \frac{1}{2}\varepsilon \, \mathbf{E}^2 + \frac{1}{2\mu_0} B^2 \tag{17}$$

but with the insertion of the extra factor $\mathrm{v}_p / \mathrm{v}_g$.

Why does group velocity enter into the energy-density expression? The Poynting vector is:

$$\begin{aligned}
\mathbf{S} &= \mu_0^{-1}\mathbf{E} \times \mathbf{B} \\
&= \mu_0^{-1}\int \!\!d\omega \, e^{-i\omega t} \boldsymbol{\mathcal{E}}(\mathbf{r},\omega) \times \int \!\!d\omega' \, e^{i\omega't} \boldsymbol{\mathcal{B}}^*(\mathbf{r},\omega')
\end{aligned} \tag{18}$$

where the $\omega'$ integral was written after changing $\omega' \to -\omega'$. For the stationary plane wave, this becomes:

$$\langle \mathbf{S} \rangle = \hat{z}\int \!\!d\omega \, I_A(\omega)\varepsilon(\omega)\mathrm{v}_p(\omega) \tag{19}$$

So, as expected, comparing Eq.(16), we observe that at each (independent) frequency:

$$(energy\ density) \times (group\ velocity) = (energy\ flux)$$

as it must be. This argument parallels those in (1, 7, 21).

In the following, we will not use the specialized Eq.(13) other than for qualitative discussions. We will use the general energy expression Eq.(7) where needed.

## 4. Inverse permittivity formulation

As discussed above, it was pointed out in (4, 8) and elsewhere, and summarized in (2) and (18), that in the context of *nonlinear* optics there is a more appropriate way to view the relation between $E$ and $D$ is. Instead of the usual Eq.(3), write:

$$\boldsymbol{\mathcal{E}}(\mathbf{r},\omega) = \eta(\mathbf{r},\omega) \cdot \boldsymbol{\mathcal{D}}(\mathbf{r},\omega) \tag{20}$$

where the inverse permittivity tensor is the matrix inverse $\eta(\mathbf{r},\omega) = [\varepsilon(\mathbf{r},\omega)]^{-1}$. Then the constitutive relation in the time domain is:

$$\mathbf{E}(\mathbf{r},t) = \int_{-\infty}^{\infty} \!\!d\omega \, e^{-i\omega t}\eta(\mathbf{r},\omega) \cdot \boldsymbol{\mathcal{D}}(\mathbf{r},\omega) \tag{21}$$

While in classical linear optics it makes no real difference which way the constitutive relation is written, in quantum nonlinear optics, where higher-order terms appear in Eq.(20), the difference is crucial, and the inverse permittivity method avoids serious errors. Therefore, we should adopt it when setting up a theory of linear optics that will be used as a foundation of a nonlinear quantum theory.

The results obtained above can be reformulated using the inverse permittivity, as:

$$W_E = \frac{1}{2} \int\limits_{-\infty}^{\infty} d\omega \int\limits_{-\infty}^{\infty} d\omega' e^{-i(\omega-\omega')t} \mathcal{D}(\mathbf{r},\omega) \cdot \left[ \frac{\omega'\eta(\mathbf{r},\omega) - \omega\eta(\mathbf{r},\omega')}{\omega' - \omega} \right] \cdot \mathcal{D}^*(\mathbf{r},\omega') \qquad (22)$$

For a stationary, incoherent plane wave in an isotropic medium:

$$\mathcal{D}(\mathbf{r},\omega) = \eta^{-1}(\mathbf{r},\omega)\hat{x}\,\mathcal{A}(\omega)e^{ikz} \qquad (23)$$

In this case the ensemble-averaged energy density is, analogously to the previous section:

$$
\begin{aligned}
\langle W_E \rangle &= \frac{1}{2} \int\limits_{-\infty}^{\infty} d\omega\, \eta^{-2}(\mathbf{r},\omega) I_A(\omega) \left\{ \lim_{\omega \to \omega'} \frac{\omega'\eta(\mathbf{r},\omega) - \omega\eta(\mathbf{r},\omega')}{\omega' - \omega} \right\} \\
&= \frac{1}{2} \int\limits_{-\infty}^{\infty} d\omega\, \eta^{-2}(\mathbf{r},\omega) I_A(\omega) \left\{ \frac{\partial}{\partial\omega'} \left[ \omega'\eta(\mathbf{r},\omega) - \omega\eta(\mathbf{r},\omega') \right] \right\}_{\omega'=\omega} \\
&= \frac{1}{2} \int\limits_{-\infty}^{\infty} d\omega\, \eta^{-2}(\mathbf{r},\omega) I_A(\omega) \left\{ \eta(\mathbf{r},\omega) - \omega\frac{\partial}{\partial\omega}\eta(\mathbf{r},\omega) \right\}
\end{aligned}
\qquad (24)
$$

And, for a nonstructured medium, it is easily shown that the ratio of phase velocity to group velocity can be expressed as:

$$\frac{v_p(\omega)}{v_g(\omega)} = \frac{1}{2\eta(\omega)} \left\{ 2\eta(\omega) - \omega\frac{\partial}{\partial\omega}\eta(\omega) \right\} \qquad (25)$$

The 'missing' factor of 2 multiplying $\eta$ inside the curly brackets in the last line of Eq.(24), which is needed to complete the expression for group velocity, comes from the magnetic energy, which is still given by Eq.(12). So the total energy density for a stationary plane wave is:

$$
\begin{aligned}
\langle W \rangle &= \frac{1}{2} \int d\omega\, I_A(\omega) \left[ \frac{1}{\eta(\omega)} + \frac{1}{\eta^2(\omega)} \left\{ \eta(\omega) - \omega\frac{\partial}{\partial\omega}\eta(\omega) \right\} \right] \\
&= \int d\omega\, I_A(\omega) \left[ \frac{1}{2\eta^2(\omega)} \left\{ 2\eta(\omega) - \omega\frac{\partial}{\partial\omega}\eta(\omega) \right\} \right] \\
&= \int d\omega\, I_A(\omega) \frac{1}{\eta(\omega)} \frac{v_p(\omega)}{v_g(\omega)}
\end{aligned}
\qquad (26)
$$

This is the same as Eq.(16), as it must be.

## 5. Mode expansions, nonorthogonality and normalization

First consider a brief aside on the mathematics of orthogonality and weight functions (Sturm-Liouville theory):

Given a differential equation in Sturm-Liouville form:

$$\partial_x^2 u(k,x) + \rho(x) k^2 u(k,x) = 0$$

Multiply by $u^*(k',x)$ and integrate:

$$\int dx\, u^*(k',x) \partial_x^2 u(k,x) + k^2 \int dx\, u^*(k',x)\rho(x)u(k,x) = 0$$

Integrate the first term by parts twice, assuming boundary terms are zero:

$$\int dx\, u(k,x) \partial_x^2 u^*(k',x) + k^2 \int dx\, u^*(k',x)\rho(x)u(k,x) = 0$$

Conjugate this whole equation and swap $k$ and $k'$, assuming $\rho(x)$ is real, to give:

$$\int dx\, u^*(k',x) \partial_x^2 u(k,x) + k^2 \int dx\, u(k,x)\rho(x)u^*(k',x) = 0$$

Subtract the first and third of these equations:

$$(k^2 - k'^2) \int dx\, u^*(k',x)\rho(x)u(k,x) = 0$$

So we see that if $k^2 \neq k'^2$ we must have the orthogonality relation:

$$\int dx\, \rho(x) u^*(k',x) u(k,x) = 0$$

where $\rho(x)$ is called a weight function. We will follow an analogous procedure below for EM modes.

Returning to our E&M problem, to quantize the field, rather than working with fields as dynamical variables directly, it is common to carry out an expansion into monochromatic spatial modes in the context of classical E&M, then quantize the amplitudes of these discrete modes. In a structured dispersive transparent medium, such as an optical fiber, a planar or ridge waveguide, or even a photonic crystal, the standard procedure is as follows (See 5, 9), following notation in (1).

Expand both $D$ and $B$ into monochromatic mode functions $\mathbf{D}_j(\mathbf{r})$ and $\mathbf{B}_j(\mathbf{r})$, which are uniquely labeled by index $j$:

$$\mathbf{D}(\mathbf{r},t) = \sum_{j=-\infty}^{\infty} \alpha_j e^{-i\omega_j t} \mathbf{D}_j(\mathbf{r}) \quad ; \quad \mathbf{B}(\mathbf{r},t) = \sum_{j=-\infty}^{\infty} \alpha_j e^{-i\omega_j t} \mathbf{B}_j(\mathbf{r}) \qquad (27)$$

The assumption that both fields can be expanded in terms of the same amplitude coefficients $\alpha_j$ is an *ansatz* that will be found to lead to a solution. To enforce that $D$ and $B$ are real, we must have $\omega_{-j} = -\omega_j$ and $\mathbf{D}_{-j}(\mathbf{r}) = \mathbf{D}_j^*(r)$, and $\mathbf{B}_{-j}(\mathbf{r}) = \mathbf{B}_j^*(r)$. The modes are as yet not normalized. In the frequency domain the expansions are:

$$\mathscr{D}(\mathbf{r},\omega) = \sum_j \alpha_j 2\pi\delta(\omega - \omega_j)\mathbf{D}_j(\mathbf{r}) \quad ; \quad \mathscr{E}(\mathbf{r},\omega) = \sum_j \alpha_j 2\pi\delta(\omega - \omega_j)\mathbf{B}_j(\mathbf{r}) \qquad (28)$$

In the case of a structured, dispersive (but not birefringent) medium, the Maxwell equations lead to the following for the modes:

$$i\omega_j \mathbf{B}_j(\mathbf{r}) = \nabla \times \left[ \eta(\mathbf{r},\omega_j)\mathbf{D}_j(\mathbf{r}) \right] \qquad (29)$$

$$-i\omega_j \mu_0 \mathbf{D}_j(\mathbf{r}) = \nabla \times \mathbf{B}_j(\mathbf{r}) \qquad (30)$$

Combining these two equations gives the eigenvalue problems:

$$\nabla \times \nabla \times \left[ \eta(\mathbf{r},\omega_j)\mathbf{D}_j(\mathbf{r}) \right] = \mu_0 \omega_j^{\ 2}\mathbf{D}_j(\mathbf{r})$$

$$\nabla \times \left[ \eta(\mathbf{r},\omega_j)\nabla \times \mathbf{B}_j(\mathbf{r}) \right] = \mu_0 \omega_j^{\ 2}\mathbf{B}_j(\mathbf{r}) \qquad (31)$$

The complication here is that $\eta(\mathbf{r},\omega_j)$ in the left-hand side of these equations depends nontrivially on the eigenvalue $\omega_j^{\ 2}$, requiring an iterative numerical solution method. A further complication is that, because each distinct solution is an eigenfunction of a different operator (via the factor $\eta(\mathbf{r},\omega_j)$), they are in general not orthogonal to one another under any scalar product integral, such as

$$\int d^3r\, \rho(\mathbf{r})\mathbf{D}_l^{\ *}(\mathbf{r}) \cdot \mathbf{D}_j(\mathbf{r}) \qquad (32)$$

with $\rho(\mathbf{r})$ being a weight function.

To find the form of an integral that defines the mode normalization (and *approximate* orthogonality), we can develop a method inspired by those in (1, 2, 7), but using *B* and *D* fields only (rather than vector potentials, dual potentials, or dynamical medium models).

Analogous to Sturm-Liouville theory, multiply Eq.(29) by $\mathbf{B}_l^{\ *}(\mathbf{r})$ and integrate over the volume *V*, to give:

$$\int d^3r \left\{ i\omega_j \mathbf{B}_j(\mathbf{r}) \cdot \mathbf{B}_l^{\ *}(\mathbf{r}) - \nabla \times \left[ \eta(\mathbf{r},\omega_j)\mathbf{D}_j(\mathbf{r}) \right] \cdot \mathbf{B}_l^{\ *}(\mathbf{r}) \right\} = 0 \qquad (33)$$

The second term can be rewritten by integrating by parts, assuming vanishing boundary values, which yields in general (2):

$$\int d^3r \left[ \nabla \times \mathbf{f}(\mathbf{r}) \right] \cdot \mathbf{g}(\mathbf{r}) = \int d^3r \left[ \nabla \times \mathbf{g}(\mathbf{r}) \right] \cdot \mathbf{f}(\mathbf{r}) \qquad (34)$$

a formula that is also a corollary of the divergence theorem of vector calculus. Thus,

$$\int d^3r \left\{ i\omega_j \mathbf{B}_j(\mathbf{r}) \cdot \mathbf{B}_l^{\ *}(\mathbf{r}) - \left[ \nabla \times \mathbf{B}_l^{\ *}(\mathbf{r}) \right] \cdot \eta(\mathbf{r},\omega_j)\mathbf{D}_j(\mathbf{r}) \right\} = 0 \qquad (35)$$

or, using the Maxwell equation Eq.(30),

$$\int d^3r \left\{ i\omega_j \mathbf{B}_j(\mathbf{r}) \cdot \mathbf{B}_l^{\ *}(\mathbf{r}) - \left[ i\omega_l \mu_0 \mathbf{D}_l^{\ *}(\mathbf{r}) \right] \cdot \eta(\mathbf{r},\omega_j)\mathbf{D}_j(\mathbf{r}) \right\} = 0 \qquad (36)$$

As in Sturm-Liouville theory, duplicate this equation, exchanging indices, ($j \leftrightarrow l$) and conjugate, giving:

$$\int d^3r \left\{ i\omega_l \mathbf{B}_l^{\ *}(\mathbf{r}) \cdot \mathbf{B}_j(\mathbf{r}) - \left[ i\omega_j \mu_0 \mathbf{D}_j(\mathbf{r}) \right] \cdot \eta(\mathbf{r},\omega_l)\mathbf{D}_l^{\ *}(\mathbf{r}) \right\} = 0 \qquad (37)$$

Subtract the two previous equations, to give:

$$\int d^3r \left\{ \left( \omega_j - \omega_l \right) \mathbf{B}_j(\mathbf{r}) \cdot \mathbf{B}_l^*(\mathbf{r}) - \left[ \omega_l \eta(\mathbf{r}, \omega_j) - \omega_j \eta(\mathbf{r}, \omega_l) \right] \mu_0 \mathbf{D}_l^*(\mathbf{r}) \cdot \mathbf{D}_j(\mathbf{r}) \right\} = 0 \quad (38)$$

Rewrite this as:

$$(\omega_j - \omega_l) \int d^3r \left\{ \mu_0^{-1} \mathbf{B}_l^*(\mathbf{r}) \cdot \mathbf{B}_j(\mathbf{r}) + \left( \frac{\omega_j \eta(\mathbf{r}, \omega_l) - \omega_l \eta(\mathbf{r}, \omega_j)}{\omega_j - \omega_l} \right) \mathbf{D}_l^*(\mathbf{r}) \cdot \mathbf{D}_j(\mathbf{r}) \right\} = 0 \quad (39)$$

For cases where $\omega_j \neq \omega_l$, the integral must equal zero:

$$\int d^3r \left\{ \mu_0^{-1} \mathbf{B}_l^*(\mathbf{r}) \cdot \mathbf{B}_j(\mathbf{r}) + \left( \frac{\omega_j \eta(\mathbf{r}, \omega_l) - \omega_l \eta(\mathbf{r}, \omega_j)}{\omega_j - \omega_l} \right) \mathbf{D}_l^*(\mathbf{r}) \cdot \mathbf{D}_j(\mathbf{r}) \right\} = 0 \quad (l \neq j) \quad (40)$$

Now, use Eq.(36) to replace the $\mathbf{B}_l^* \cdot \mathbf{B}_j$ term in Eq.(40), giving:

$$\int d^3r \, \mathbf{D}_l^*(\mathbf{r}) \cdot \mathbf{D}_j(\mathbf{r}) \left\{ \frac{\omega_l}{\omega_j} \eta(\mathbf{r}, \omega_j) + \frac{\omega_j \eta(\mathbf{r}, \omega_l) - \omega_l \eta(\mathbf{r}, \omega_j)}{\omega_j - \omega_l} \right\} = 0 \quad (l \neq j) \quad (41)$$

This result *resembles* an orthogonality integral for the displacement field, but the term in brackets is not a general-purpose weight function because it depends on $\omega_j$ and $\omega_l$. Equation (41) seems not to have appeared in previous publications. Note that if $\eta(\mathbf{r}, \omega_j)$ is independent of frequency, becoming $\eta(\mathbf{r})$, the left-hand side of Eq.(41) reduces to a form identical to the familiar, expected Eq.(32): $\int d^3r \, \eta(\mathbf{r}) \mathbf{D}_l^*(\mathbf{r}) \cdot \mathbf{D}_j(\mathbf{r}) = 0 \ (l \neq j)$

The fact that the modes are not orthogonal under a general weight function is a result of dispersion (1). This means Eq.(41) cannot be used to project an arbitrary mode from a given field distribution $\mathbf{D}(\mathbf{r}, t)$. The above derivation generalizes the Sturm-Liouville-inspired derivation presented in the case of one-dimensional propagation by Santos and Loudon (25).

To determine a **normalization condition** for the modes, note that for $\omega_j = \omega_l$ the integral in Eq.(39) has a nonzero value, which we denote as $M_j$ (its value is arbitrary and we specify it later):

$$\int d^3r \left\{ \mu_0^{-1} \mathbf{B}_j^*(\mathbf{r}) \cdot \mathbf{B}_j(\mathbf{r}) + \lim_{\omega \to \omega_j} \left( \frac{\omega_j \eta(\mathbf{r}, \omega) - \omega \eta(\mathbf{r}, \omega_j)}{\omega_j - \omega} \right) \mathbf{D}_j^*(\mathbf{r}) \cdot \mathbf{D}_j(\mathbf{r}) \right\} \doteq M_j \quad (42)$$

Again, we can use Eq.(36) to replace the $\mathbf{B}_l^* \cdot \mathbf{B}_j$ term, giving:

$$\int d^3r \, \mathbf{D}_j^*(\mathbf{r}) \cdot \mathbf{D}_j(\mathbf{r}) \lim_{\omega \to \omega_j} \left\{ \frac{\omega}{\omega_j} \eta(\mathbf{r}, \omega_j) + \frac{\omega_j \eta(\mathbf{r}, \omega) - \omega \eta(\mathbf{r}, \omega_j)}{\omega_j - \omega} \right\} = M_j \quad (43)$$

Similar to the calculations in Eq.(24), the limit of the term in brackets becomes, generalizing Eq.(25):

$$\lim_{\omega \to \omega_j} \left\{ \frac{\omega}{\omega_j} \eta(\mathbf{r}, \omega_j) + \frac{\omega_j \eta(\mathbf{r}, \omega) - \omega \eta(\mathbf{r}, \omega_j)}{\omega_j - \omega} \right\}$$

$$= 2 \left\{ \eta(\mathbf{r}, \omega) - \frac{\omega}{2} \frac{\partial}{\partial \omega} \eta(\mathbf{r}, \omega) \right\}_{\omega_j} = 2 \eta(\mathbf{r}, \omega) \frac{v_p(\mathbf{r}, \omega)}{v_g(\mathbf{r}, \omega)} \tag{44}$$

This gives the normalization integral:

$$\int d^3 r \, \eta(\mathbf{r}, \omega_j) R(\mathbf{r}, \omega_j) \mathbf{D}_j^*(\mathbf{r}) \cdot \mathbf{D}_j(\mathbf{r}) = \frac{1}{2} M_j \tag{45}$$

where the *ratio* of 'local' phase velocity $v_p(\mathbf{r}, \omega)$ to group velocity $v_g(\mathbf{r}, \omega)$ is denoted $R(\mathbf{r}, \omega)$, and takes on various forms:

$$R(\mathbf{r}, \omega) \equiv \frac{v_p(\mathbf{r}, \omega)}{v_g(\mathbf{r}, \omega)} = \frac{1}{2\eta(\mathbf{r}, \omega)} \left\{ 2\eta(\mathbf{r}, \omega) - \omega \frac{\partial}{\partial \omega} \eta(\mathbf{r}, \omega) \right\}$$

$$= 1 - \frac{\omega}{2\eta(\mathbf{r}, \omega)} \frac{\partial}{\partial \omega} \eta(\mathbf{r}, \omega) = 1 - \frac{\omega}{2} n^2(\mathbf{r}, \omega) \frac{\partial}{\partial \omega} \frac{1}{n^2(\mathbf{r}, \omega)} \ . \tag{46}$$

$$= 1 + \frac{\omega}{n(\mathbf{r}, \omega)} \frac{\partial}{\partial \omega} n(\mathbf{r}, \omega) = \frac{1}{n(\mathbf{r}, \omega)} \frac{\partial}{\partial \omega} \left( \omega \, n(\mathbf{r}, \omega) \right)$$

where we introduced the local refractive index $n^2(\mathbf{r}, \omega)$, such that $\varepsilon(\mathbf{r}, \omega) = \varepsilon_0 n^2(\mathbf{r}, \omega)$, or $\eta(\mathbf{r}, \omega) = 1 / \left( \varepsilon_0 n^2(\mathbf{r}, \omega) \right)$. Note that 'local velocities' are not really physical velocities, but are a convenient definition for describing local material properties. The recognition that the extra factor appearing in the normalization integral (which has been long known) equals the ratio of velocities is due to Sipe and reported in (1), and Drummond (7) noted a similar relation.

## 6. Hamiltonian and Quantization

To quantize the field, we need to have, at minimum, a suitable Hamiltonian and appropriate pairs of conjugate variables. The latter can be determined using Lagrangian methods, but here we will simply postulate them and verify that using Hamilton's equations generates the correct Maxwell equations. This less formal approach is simpler for the task at hand, although not as powerful as the Lagrangian method. It is in the spirit of the treatment in (11), but developed here for the case of structured media, using the mode quasi-orthogonality formulas derived above.

While Eq.(7) gives a spatially local expression for the electric energy density (at location $\mathbf{r}$), recall that it is 'nonlocal' in frequency, requiring a double integral over that variable. The double integral reduces to a single integral only if the field is monochromatic or the field components are statistically uncorrelated, as in Eq.(12). Yet, to quantize a system we need only the total energy, not a local energy density. To this purpose, Loudon pointed out that even in the presence of dispersion it is possible to express the total energy integrated over the entire volume as a single frequency integral (26). Loudon's method was generalized to include spatial structure of the dielectric in (16).

Therefore, for the purpose of quantization we define a Hamiltonian for the macroscopic system as the integral of the energy density in the relevant volume $V$:

$$\mathscr{H} = \mathscr{H}_B + \mathscr{H}_E = \int_V d^3r \left\{ W_B + W_E \right\} \qquad (47)$$

Using the frequency-domain version, Eq.(28), of the mode expansion in Eq.(22) gives:

$$\mathscr{H}_E = \frac{1}{2} \int_V d^3r \int_{-\infty}^{\infty} d\omega \int_{-\infty}^{\infty} d\omega' \, e^{-i(\omega-\omega')t} \times$$

$$\sum_j \alpha_j 2\pi\delta(\omega-\omega_j) \mathbf{D}_j(\mathbf{r}) \cdot \left[ \frac{\omega'\eta(\mathbf{r},\omega) - \omega\eta(\mathbf{r},\omega')}{\omega'-\omega} \right] \cdot \sum_l \alpha_l^* 2\pi\delta(\omega'-\omega_l) \mathbf{D}_l^*(\mathbf{r}) \quad (48)$$

$$= \frac{1}{2} \sum_j \alpha_j \sum_l \alpha_l^* e^{-i(\omega_j-\omega_l)t} \int_V d^3r \, \mathbf{D}_j(\mathbf{r}) \cdot \left[ \frac{\omega_l\eta(\mathbf{r},\omega_j) - \omega_j\eta(\mathbf{r},\omega_l)}{\omega_l-\omega_j} \right] \cdot \mathbf{D}_l^*(\mathbf{r})$$

and, likewise, for the magnetic energy:

$$\mathscr{H}_B = \frac{1}{2\mu_0} \sum_j \alpha_j \sum_l \alpha_l^* e^{-i(\omega_j-\omega_l)t} \int_V d^3r \, \mathbf{B}_j(\mathbf{r}) \cdot \mathbf{B}_l^*(\mathbf{r}) \qquad (49)$$

Combining these gives:

$$\mathscr{H} = \sum_j \alpha_j \sum_l \alpha_l^* e^{-i(\omega_j-\omega_l)t} \times$$

$$\frac{1}{2} \int_V d^3r \left\{ \frac{1}{\mu_0} \mathbf{B}_j(\mathbf{r}) \cdot \mathbf{B}_l^*(\mathbf{r}) + \mathbf{D}_j(\mathbf{r}) \cdot \left( \frac{\omega_j\eta(\mathbf{r},\omega_l) - \omega_l\eta(\mathbf{r},\omega_j)}{\omega_j-\omega_l} \right) \cdot \mathbf{D}_l^*(\mathbf{r}) \right\} \qquad (50)$$

We can recognize the integral in this equation as none other than the quasi-orthogonality integral in Eq.(39) (or Eq.(41)). Therefore, the double sum reduces to a single sum, which we write as:

$$\mathscr{H} = \sum_{j=-\infty}^{\infty} \frac{1}{2} M_j \alpha_j^* \alpha_j$$

$$= \sum_{j=0}^{\infty} \frac{1}{2} M_j \left( \alpha_j^* \alpha_j + \alpha_j \alpha_j^* \right) \qquad (51)$$

Now, choose $M_j = \hbar\omega_j$, and define real variables (quadrature amplitudes) by:

$$\alpha_j = \frac{1}{\sqrt{2\hbar}} \left( Q_j + iP_j \right)$$

so that

$$\mathcal{H} = \frac{1}{2}\sum_{j=0}^{\infty}\omega_j\left(Q_j^{\ 2} + P_j^{\ 2}\right) \tag{52}$$

The total energy of the system has thus been expressed as a single sum over frequency, which in a continuum limit can be replaced by an integral over frequency, consistent with other familiar treatments. The total energy is thus 'local' in frequency, but not expressible as the integral of an energy density that is local in both space and frequency, as discussed at the beginning of this section. This form of the energy is made possible by expanding using the appropriate spatial modes and summing over the mode index $j$, which is the true meaning of Eq.(52).

This Hamiltonian, along with the Hamilton equations, $\dot{Q}_j = \partial\mathcal{H}/\partial P_j$, $\dot{P}_j = -\partial\mathcal{H}/\partial Q_j$ generate the equations of motion $\dot{\alpha}_j = -i\omega_j\alpha_j$, which agrees with the classical forms in Eq.(27). Therefore, to quantize this theory, we can simply elevate the variables to operators and invoke the commutation relations $[\hat{Q}_j, \hat{P}_k] = i\hbar\delta_{jk}$. This leads to the bosonic creation and annihilation operators defined as $\hat{a}_j = (\hat{Q}_j + i\hat{P}_j)/\sqrt{2\hbar}$, with

$$\left[\hat{a}_j(t), \hat{a}_l^\dagger(t)\right] = \delta_{jl} \tag{53}$$

Then

$$\mathcal{H} = \sum_{j=0}^{\infty}\frac{1}{2}\hbar\omega_j\left(\hat{a}_j^\dagger\hat{a}_j + \hat{a}_j\hat{a}_j^\dagger\right) = \sum_{j=0}^{\infty}\hbar\omega_j\left(\hat{a}_j^\dagger\hat{a}_j + \frac{1}{2}\right) \tag{54}$$

This choice of mode normalization, $M_j = \hbar\omega_j$, results in the final form for the mode normalization in the quantum theory:

$$\int d^3r\,\eta_j(\mathbf{r})R_j(\mathbf{r})\mathbf{D}_j^{\ *}(\mathbf{r})\cdot\mathbf{D}_j(\mathbf{r}) = \frac{\hbar\omega_j}{2} \tag{55}$$

where we introduced the abbreviations $\eta_j(\mathbf{r}) \equiv \eta(\mathbf{r}, \omega_j)$, $R_j(\mathbf{r}) \equiv R(\mathbf{r}, \omega_j)$. This normalization integral is in agreement with Eq.(32) of (1), which resulted from using a dynamical model for the medium.

Note that if the medium is not dispersive, so the 'local' group and phase velocities are equal ($\mathrm{v}_{gj}(\mathbf{r}) = \mathrm{v}_{pj}(\mathbf{r})$), so $R_j(\mathbf{r}) = 1$, this integral is proportional to that conjectured in Eq.(32).

We now introduce new normalized modes in a form more familiar in quantum optics theory. Define new modes $\mathbf{u}_j(\mathbf{r})$ by multiplying with position-independent factors, so the new modes still obey the Maxwell equations:

$$\mathbf{D}_j(\mathbf{r}) \equiv i\sqrt{\frac{\varepsilon_0\hbar\omega_j}{2}}\,\mathbf{u}_j(\mathbf{r}) \tag{56}$$

where $i$ is inserted for consistency with a standard phase convention in quantum optics theory. Then the normalization integral for the displacement-field modes becomes:

$$\int d^3r\,\varepsilon_0\eta_j(\mathbf{r})R_j(\mathbf{r})\mathbf{u}_j^{\ *}(\mathbf{r})\cdot\mathbf{u}_j(\mathbf{r}) = 1 \tag{57}$$

where now only medium properties enter the normalization. With these modes, the positive-frequency part of the displacement-field operator is:

$$\hat{\mathbf{D}}^{(+)}(\mathbf{r},t) = i\sum_{j=0}^{\infty}\sqrt{\frac{\varepsilon_0 \hbar \omega_j}{2}}\,\hat{a}_j e^{-i\omega_j t}\,\mathbf{u}_j(\mathbf{r}) \tag{58}$$

whereas the negative-frequency part is $\hat{\mathbf{D}}^{(-)} = \hat{\mathbf{D}}^{(+)\dagger}$. The form of the magnetic-field modes is found using the Maxwell equation for the modes and Eq.(56):

$$\begin{aligned}\mathbf{B}_j(\mathbf{r}) &= (i\omega_j)^{-1}\nabla\times\left[\eta_j(\mathbf{r})\mathbf{D}_j(\mathbf{r})\right] \\ &= (i\omega_j)^{-1}i\sqrt{\frac{\hbar\omega_j}{2\varepsilon_0}}\,\nabla\times\left[\varepsilon_0\eta_j(\mathbf{r})\mathbf{u}_j(\mathbf{r})\right]\end{aligned} \tag{59}$$

So, Eq.(27) gives:

$$\begin{aligned}\hat{\mathbf{B}}^{(+)}(\mathbf{r},t) &= \frac{i}{c}\sum_{j=-\infty}^{\infty}\sqrt{\frac{\hbar\omega_j}{2\varepsilon_0}}\,\hat{a}_j e^{-i\omega_j t}\frac{c}{i\omega_j}\nabla\times\left[\varepsilon_0\eta_j(\mathbf{r})\mathbf{u}_j(\mathbf{r})\right] \\ &\doteq \frac{i}{c}\sum_{j=-\infty}^{\infty}\sqrt{\frac{\hbar\omega_j}{2\varepsilon_0}}\,\hat{a}_j e^{-i\omega_j t}\tilde{\mathbf{u}}_j(\mathbf{r})\end{aligned} \tag{60}$$

where the new magnetic modes are:

$$\begin{aligned}\tilde{\mathbf{u}}_j(\mathbf{r}) &= \frac{c}{i\omega_j}\nabla\times\left[\varepsilon_0\eta_j(\mathbf{r})\mathbf{u}_j(\mathbf{r})\right] \\ &= \frac{1}{ic\mu_0\omega_j}\nabla\times\left[\eta_j(\mathbf{r})\mathbf{u}_j(\mathbf{r})\right]\end{aligned} \tag{61}$$

Comparing above formulas, the unscaled and scaled magnetic modes are related by:

$$\mathbf{B}_j(\mathbf{r})\doteq\frac{i}{c}\sqrt{\frac{\hbar\omega_j}{2\varepsilon_0}}\,\tilde{\mathbf{u}}_j(\mathbf{r}) \tag{62}$$

and $\hat{\mathbf{B}}^{(-)} = \hat{\mathbf{B}}^{(+)\dagger}$. And, using Eq.(30), the inverse relation between electric and magnetic modes is

$$\mathbf{u}_j(\mathbf{r}) = \frac{ic}{\omega_j}\nabla\times\tilde{\mathbf{u}}_j(\mathbf{r})$$

The electric field, using $\mathcal{E}(\underline{r},\omega) = \eta(\underline{r},\omega)\cdot\mathcal{D}(\underline{r},\omega)$, is

$$\hat{\mathbf{E}}^{(+)}(\mathbf{r},t) = i\sum_{j=0}^{\infty}\sqrt{\frac{\hbar\omega_j}{2\varepsilon_0}}\hat{a}_j e^{-i\omega_j t}\varepsilon_0\eta_j(\mathbf{r})\mathbf{u}_j(\mathbf{r}) \qquad (63)$$

## 7. Fields in a dispersive, nonstructured medium

In a nonstructured (and non-birefringent) medium, the normalization integral Eq.(57) becomes:

$$\int d^3 r\,\mathbf{u}_j^{*}(\mathbf{r})\cdot\mathbf{u}_j(\mathbf{r}) = \frac{1}{\varepsilon_0\eta_j R_j} \qquad (64)$$

The fields and modes simplify to:

$$\hat{\mathbf{D}}^{(+)}(\mathbf{r},t) = i\sum_{j=0}^{\infty}\sqrt{\frac{\varepsilon_0\hbar\omega_j}{2}}\hat{a}_j e^{-i\omega_j t}\,\mathbf{u}_j(\mathbf{r})$$

$$\hat{\mathbf{B}}^{(+)}(\mathbf{r},t) = \frac{i}{c}\sum_{j=-\infty}^{\infty}\sqrt{\frac{\hbar\omega_j}{2\varepsilon_0}}\hat{a}_j e^{-i\omega_j t}\,\tilde{\mathbf{u}}_j(\mathbf{r})$$

$$\tilde{\mathbf{u}}_j(\mathbf{r}) = \frac{c\varepsilon_0\eta_j}{i\omega_j}\nabla\times\mathbf{u}_j(\mathbf{r}) \qquad (65)$$

$$\hat{\mathbf{E}}^{(+)}(\mathbf{r},t) = i\sum_{j=0}^{\infty}\sqrt{\frac{\hbar\omega_j\varepsilon_0\eta_j^{2}}{2}}\hat{a}_j e^{-i\omega_j t}\,\mathbf{u}_j(\mathbf{r})$$

## 8. Plane waves in a dispersive, nonstructured medium

For linearly polarized plane waves in a dispersive, nonstructured medium, the mode normalization Eq.(64) implies:

$$\mathbf{u}_j(\mathbf{r}) = \sqrt{\frac{1}{\varepsilon_0\eta_j R_j}}\,\mathbf{e}_j\frac{e^{i\mathbf{k}_j\cdot\mathbf{r}}}{\sqrt{V}} \qquad (66)$$

with $\mathbf{k}_j = k_j\hat{\mathbf{k}}_j$, where $\hat{\mathbf{k}}$ is a unit vector in the propagation direction, and

$$\left|\mathbf{k}_j\right| = k_j = \omega_j n(\omega_j)/c = (\omega_j/c)\sqrt{\varepsilon(\omega_j)/\varepsilon_0} \qquad (67)$$

So, the fields become, from Eq.(58):

$$\hat{\mathbf{D}}^{(+)}(\mathbf{r},t) = i\sum_{j=0}^{\infty}\sqrt{\frac{\varepsilon_0\hbar\omega_j}{2}}\,\hat{a}_j e^{-i\omega_j t}\sqrt{\frac{1}{\varepsilon_0\eta_j R_j}}\mathbf{e}_j\frac{e^{i\mathbf{k}_j\bullet\mathbf{r}}}{\sqrt{V}}$$

$$\hat{\mathbf{E}}^{(+)}(\mathbf{r},t) = i\sum_{j=0}^{\infty}\sqrt{\frac{\hbar\omega_j}{2\varepsilon_0}}\,\hat{a}_j e^{-i\omega_j t}\sqrt{\frac{\varepsilon_0\eta_j}{R_j}}\mathbf{e}_j\frac{e^{i\mathbf{k}_j\bullet\mathbf{r}}}{\sqrt{V}}$$

(68)

or, using $\mathrm{v}_p = c/n_j$ and $\varepsilon_0\eta_j = 1/n_j{}^2$:

$$\hat{\mathbf{D}}^{(+)}(\mathbf{r},t) = i\sum_{j=0}^{\infty}\sqrt{\frac{\varepsilon_0\hbar\omega_j}{2}\frac{\mathrm{v}_{gj}n_j{}^3}{c}}\,\hat{a}_j e^{-i\omega_j t}\,\mathbf{e}_j\frac{e^{i\mathbf{k}_j\bullet\mathbf{r}}}{\sqrt{V}}$$

$$\hat{\mathbf{E}}^{(+)}(\mathbf{r},t) = i\sum_{j=0}^{\infty}\sqrt{\frac{\hbar\omega_j}{2\varepsilon_0}\frac{\mathrm{v}_{gj}}{n_j c}}\,\hat{a}_j e^{-i\omega_j t}\mathbf{e}_j\frac{e^{i\mathbf{k}_j\bullet\mathbf{r}}}{\sqrt{V}}$$

(69)

This form agrees with references (1) and with (11), accounting for different unit systems.

The magnetic-field modes become:

$$\tilde{\mathbf{u}}_j(\mathbf{r}) = \frac{c\varepsilon_0\eta_j}{i\omega_j}\nabla\times\mathbf{u}_j(\mathbf{r}) = \frac{c\varepsilon_0\eta_j}{i\omega_j}\nabla\times\mathbf{e}_j\sqrt{\frac{1}{\varepsilon_0\eta_j R_j}}\frac{e^{i\mathbf{k}_j\bullet\mathbf{r}}}{\sqrt{V}}$$

$$= \sqrt{\frac{\varepsilon_0\eta_j}{R_j}}\frac{c}{i\omega_j}ik_j\left(\hat{\mathbf{k}}_j\times\mathbf{e}_j\right)\frac{e^{i\mathbf{k}_j\bullet\mathbf{r}}}{\sqrt{V}}$$

(70)

So the magnetic field becomes:

$$\hat{\mathbf{B}}^{(+)}(\mathbf{r},t) = \frac{i}{c}\sum_{j=0}^{\infty}\sqrt{\frac{\hbar\omega_j}{2\varepsilon_0}}\,\hat{a}_j e^{-i\omega_j t}\sqrt{\frac{\varepsilon_0\eta_j}{R_j}}\frac{c}{i\omega_j}ik_j\left(\hat{\mathbf{k}}_j\times\mathbf{e}_j\right)\frac{e^{i\mathbf{k}_j\bullet\mathbf{r}}}{\sqrt{V}}$$

$$= \frac{i}{c}\sum_{j=0}^{\infty}\sqrt{\frac{\hbar\omega_j}{2\varepsilon_0}}\,\hat{a}_j e^{-i\omega_j t}\sqrt{\frac{\varepsilon_0\eta_j}{R_j}}n_j\left(\hat{\mathbf{k}}_j\times\mathbf{e}_j\right)\frac{e^{i\mathbf{k}_j\bullet\mathbf{r}}}{\sqrt{V}}$$

(71)

## 9. Group velocity and Energy Flux

We see the group velocity appearing in the quantum-field-operator coefficients in Eq.(69). To understand this fact, note that the quantum counterpart of Eq.(18), is the Hermetian-operator version of the Poynting vector,

$$\mathbf{S} = \frac{1}{2}\mu_0^{-1}\hat{\mathbf{E}}\times\hat{\mathbf{B}} - \frac{1}{2}\mu_0^{-1}\hat{\mathbf{B}}\times\hat{\mathbf{E}}$$

(72)

with the minus sign arising from the nature of the cross product. Using $\hat{\mathbf{E}} = \hat{\mathbf{E}}^{(-)} + \hat{\mathbf{E}}^{(+)}$ and likewise for $\hat{\mathbf{B}}$, we obtain eight terms. Time averaging over a few optical cycles leaves four terms:

$$\overline{\mathbf{S}} = \frac{1}{2}\mu_0^{-1}\hat{\mathbf{E}}^{(-)}\times\hat{\mathbf{B}}^{(+)} - \frac{1}{2}\mu_0^{-1}\hat{\mathbf{B}}^{(-)}\times\hat{\mathbf{E}}^{(+)}$$
$$+\frac{1}{2}\mu_0^{-1}\hat{\mathbf{E}}^{(+)}\times\hat{\mathbf{B}}^{(-)} - \frac{1}{2}\mu_0^{-1}\hat{\mathbf{B}}^{(+)}\times\hat{\mathbf{E}}^{(-)} \tag{73}$$

Then for a *single* monochromatic plane-wave mode $j$ in a dispersive nonstructured medium, using $\mathrm{v}_{pj}=c/n_j$, $\varepsilon_0\eta_j=1/n_j^2$, $\varepsilon_j=\varepsilon_0 n_j^2$, $\varepsilon_0\mu_0 c^2=1$, $1/R_j=\mathrm{v}_{gj}/\mathrm{v}_{pj}$, we find the first term is

$$\frac{1}{2}\mu_0^{-1}\hat{\mathbf{E}}^{(-)}\times\hat{\mathbf{B}}^{(+)} = \frac{1}{2\mu_0}\sqrt{\frac{\hbar\omega_j}{2\varepsilon_0}}\hat{a}_j^\dagger e^{i\omega_j t}\sqrt{\frac{\varepsilon_0\eta_j}{R_j}}\mathbf{e}_j\frac{e^{-i\mathbf{k}_j\cdot\mathbf{r}}}{\sqrt{V}}\times\frac{1}{c}\sqrt{\frac{\hbar\omega_j}{2\varepsilon_0}}\hat{a}_j e^{-i\omega_j t}\sqrt{\frac{\varepsilon_0\eta_j}{R_j}}\,n_j\left(\hat{\mathbf{k}}_j\times\mathbf{e}_j\right)\frac{e^{i\mathbf{k}_j\cdot\mathbf{r}}}{\sqrt{V}}$$
$$=\frac{1}{2\mu_0}\hat{\mathbf{k}}_j\hat{a}_j^\dagger\hat{a}_j\frac{\hbar\omega_j}{2c\varepsilon_0}\frac{\varepsilon_0\eta_j\mathrm{v}_{gj}}{\mathrm{v}_{pj}}\frac{n_j}{V} \tag{74}$$
$$=\frac{1}{2}\hat{\mathbf{k}}_j\hat{a}_j^\dagger\hat{a}_j\frac{\hbar\omega_j}{2V}\mathrm{v}_{gj}$$

The third term in Eq.(73) contributes a similar piece, with $\hat{a}_j^\dagger\hat{a}_j$ replaced by $\hat{a}_j\hat{a}_j^\dagger = \hat{a}_j^\dagger\hat{a}_j + 1$. The second and fourth terms contribute the same results as the first and third. Then, dropping the terms corresponding to the (infinite) vacuum energy leaves, finally,

$$\overline{\mathbf{S}} = \hat{\mathbf{k}}_j\hat{a}_j^\dagger\hat{a}_j\frac{\hbar\omega_j}{V}\mathrm{v}_{gj} \tag{75}$$

This dependence on energy density and group velocity is as expected for a monochromatic field, consistent with ideas discussed earlier in Section 3. If the field is broadband with correlations between mode amplitudes, cross terms between mode operators representing different frequencies will appear.

## 10. Plane-wave fields in vacuum

The electric field in vacuum is:

$$\hat{\mathbf{E}}^{(+)}(\mathbf{r},t) = i\sum_{j=0}^\infty\sqrt{\frac{\hbar\omega_j}{2\varepsilon_0}}\hat{a}_j e^{-i\omega_j t}\mathbf{e}_j\frac{e^{i\mathbf{k}_j\cdot\mathbf{r}}}{\sqrt{V}} \tag{76}$$

in agreement with the standard form (27). For the magnetic field:

$$\tilde{\mathbf{u}}_j(\mathbf{r}) = \frac{\nabla\times\mathbf{u}_j(\mathbf{r})}{ik_j} = \frac{\nabla\times\mathbf{e}_j e^{i\mathbf{k}_j\cdot\mathbf{r}}}{ik_j\sqrt{V}} = (\hat{\mathbf{k}}_j\times\mathbf{e}_j)\frac{e^{i\mathbf{k}_j\cdot\mathbf{r}}}{\sqrt{V}}$$
$$\hat{\mathbf{B}}^{(+)}(\mathbf{r},t) = \frac{i}{c}\sum_{j=0}^\infty\sqrt{\frac{\hbar\omega_j}{2\varepsilon_0}}\hat{a}_j e^{-i\omega_j t}(\hat{\mathbf{k}}_j\times\mathbf{e}_j)\frac{e^{i\mathbf{k}_j\cdot\mathbf{r}}}{\sqrt{V}} \tag{77}$$

## 11. Band Approximation and Continuum Limit

Here we assume that the fields propagating in the medium are in distinct frequency bands, denoted as $B_J$, centered at well-separated center frequencies $\bar{\omega}_J$, with $J = 1, 2, 3, \ldots$. Consider plane waves traveling in the $z$ direction in a homogeneous, time-stationary, but dispersive, medium. Denote the mode indices falling within the band by $j \in J$. From Eq.(69), the field in the $B_J$ band is then:

$$\hat{\mathbf{D}}^{(+)}(\mathbf{r},t)_{Band\,J} = i \sum_{j \in J} \sqrt{\frac{\varepsilon_0 \hbar \omega_j \, \mathrm{v}_{gj} n_j^{\,3}}{2}} \, \hat{a}_j e^{-i\omega_j t} \, \mathbf{e}_j \frac{e^{i\beta_j z}}{\sqrt{AL}}$$

$$\hat{\mathbf{E}}^{(+)}(\mathbf{r},t)_{Band\,J} = i \sum_{j \in J} \sqrt{\frac{\hbar \omega_j \, \mathrm{v}_{gj}}{2\varepsilon_0 \, n_j c}} \, \hat{a}_j e^{-i\omega_j t} \mathbf{e}_j \frac{e^{i\beta_j z}}{\sqrt{AL}}$$

(78)

where $A$ and $L$ are the transverse area and length of the quantization volume, and $\beta_j$ is the propagation constant. Periodic boundary conditions require:

$$\exp[i\beta_j(z+L)] = \exp[i\beta_j z] \tag{79}$$

so $\beta_j = j 2\pi / L$. Note this condition is valid even in the presence of dispersion because $\beta_j$ is simply $2\pi$ times the inverse wavelength of a spatially periodic wave. (Although the corresponding mode frequencies are not uniformly spaced.)

In the continuum limit, $L \to \infty$, we replace $\beta_j \to \beta$, $\omega_j \to \omega(\beta)$, and $\Sigma_j \to L/2\pi \int d\beta$. Then

$$\left[\hat{a}_j, \hat{a}_{j'}^\dagger\right] = \delta_{jj'} \;\Rightarrow\; \frac{L}{2\pi} \int_{-\infty}^{\infty} d\beta \left[\hat{a}(\beta), \hat{a}^\dagger(\beta')\right] = 1 \tag{80}$$

Choose: $\hat{a}_j \doteq \sqrt{L^{-1}}\, \hat{b}(\beta)$, so $[\hat{b}(\beta), \hat{b}^\dagger(\beta')] = 2\pi\delta(\beta - \beta')$. Then the displacement field operator becomes (using $B_J$ to label the band):

$$\hat{\mathbf{D}}^{(+)}(\mathbf{r},t)_{Band\,J} = i \frac{L}{2\pi} \int_{B_J} d\beta \sqrt{\frac{\varepsilon_0 \hbar \omega(\beta) \, \mathrm{v}_{gJ} n_J^{\,3}}{2}} \sqrt{L^{-1}} \, \hat{b}(\beta) e^{-i\omega(\beta)t} \, \mathbf{e}_J \frac{\exp[i\beta z]}{\sqrt{AL}}$$

$$= i \int_{B_J} \frac{d\beta}{2\pi} \sqrt{\frac{\varepsilon_0 \hbar \omega(\beta)}{2} \frac{\mathrm{v}_{gJ} n_J^{\,3}}{c}} \, \hat{b}(\beta) e^{-i\omega(\beta)t} \, \mathbf{e}_J \frac{\exp[i\beta z]}{\sqrt{A}}$$

(81)

And the electric field becomes:

$$\hat{\mathbf{E}}^{(+)}(\mathbf{r},t)_{Band\ J} = i \int_{B_J} \frac{d\beta}{2\pi} \sqrt{\frac{\hbar\omega(\beta)}{2\varepsilon_0} \frac{\mathrm{v}_{gJ}}{n_J c}} \hat{b}(\beta) e^{-i\omega(\beta)t} \ \mathbf{e}_J \frac{\exp[i\beta z]}{\sqrt{A}}$$

(82)

Note that $L$ no longer appears.

## 12. Frequency labeling of modes and operators

When propagation is one dimensional, as in a waveguide, it is often useful to adopt angular frequency $\omega$ as the continuous variable labeling modes, since we typically measure frequency. Recall for each transverse mode (see below) there is a dispersion relation $\omega(\beta)$, determined by solutions of the mode eigenvalue equation. Changing variables to frequency and rescaling the operators:

$$d\beta = \frac{d\beta}{d\omega} d\omega = \frac{1}{\mathrm{v}_g} d\omega \quad , \quad \hat{c}(\omega) \doteq \sqrt{\frac{d\beta}{d\omega}} \hat{b}(\beta) = \sqrt{\frac{1}{\mathrm{v}_g}} \hat{b}(\beta) \ , \ \hat{b}(\beta) = \sqrt{\mathrm{v}_g} \hat{c}(\omega)$$

(83)

$$\left[ \hat{b}(\beta), \hat{b}^\dagger(\beta') \right] = 2\pi\delta(\beta-\beta') \Rightarrow \left[ \hat{c}(\omega), \hat{c}^\dagger(\omega') \right] = \frac{d\beta}{d\omega} 2\pi\delta(\beta-\beta') = 2\pi\delta(\omega-\omega')$$

Then:

$$\hat{\mathbf{D}}^{(+)}(\underline{r},t)_{Band\ J} = i \int_{B_J} \left( \frac{d\omega}{2\pi} \frac{1}{\mathrm{v}_{gJ}} \right) \sqrt{\frac{\varepsilon_0 \hbar\omega}{2} \frac{\mathrm{v}_{gJ} n_J{}^3}{c}} \sqrt{\mathrm{v}_g} \hat{c}(\omega) e^{-i\omega t} \ \mathbf{e}_J \frac{\exp[i\beta(\omega)z]}{\sqrt{A}}$$

$$= i \int_{B_J} \frac{d\omega}{2\pi} \sqrt{\frac{\varepsilon_0 \hbar\omega}{2} \frac{n_J{}^3}{c}} \ \hat{c}(\omega) e^{-i\omega t} \ \mathbf{e}_J \frac{\exp[i\beta(\omega)z]}{\sqrt{A}}$$

$$\hat{\mathbf{E}}^{(+)}(\underline{r},t)_{Band\ J} = i \int_{B_J} \frac{d\omega}{2\pi} \sqrt{\frac{\hbar\omega}{2\varepsilon_0 n_J c}} \ \hat{c}(\omega) e^{-i\omega t} \ \mathbf{e}_J \frac{\exp[i\beta(\omega)z]}{\sqrt{A}}$$

(84)

We see that the group velocity does not appear here, although refractive index appears in a nonobvious way.

If nontrivial dynamics occur, the annihilation operators become time-dependent, $\hat{c}(\omega,t)$. It seems slightly strange to have both time and frequency as arguments of the annihilation operators, but recall that here frequency $\omega$ really labels $\beta$, which is $2\pi$ times the spatial period (inverse wave length); it is not the Fourier transform variable conjugate to time.

## 13. Approximate Mode Projector

Returning to the case of spatially structured media, we show that in a weakly dispersive medium, as is common in transparent media, we can find an approximate procedure to project a $\mathbf{D}_j(\mathbf{r})$ mode from the total field. In this case, the term in brackets in Eq.(41) can be approximated as

$$\approx \left\{ \eta(\mathbf{r},\omega_j) + \frac{\omega_j \eta(\mathbf{r},\omega_l) - \omega_l \eta(\mathbf{r},\omega_j)}{\omega_j - \omega_l} \right\} \approx \left\{ 2\eta(\mathbf{r},\omega_j) - \omega_j \frac{\eta(\mathbf{r},\omega_j) - \eta(\mathbf{r},\omega_l)}{\omega_j - \omega_l} \right\} \approx$$

$$\left\{ 2\eta(\mathbf{r},\omega_j) - \omega_j \frac{\partial}{\partial \omega_j} \eta(\mathbf{r},\omega_j) \right\} = 2\eta(\mathbf{r},\omega_j) R(\mathbf{r},\omega_j) \approx 2\eta(\mathbf{r},\omega_l) R(\mathbf{r},\omega_l) \tag{85}$$

Thus, in this approximation,

$$\int d^3 r\, \mathbf{D}_l^*(\mathbf{r}) \cdot \mathbf{D}_j(\mathbf{r}) \eta(\mathbf{r},\omega_l) R(\mathbf{r},\omega_j) \approx 0 \quad (l \neq j) \tag{86}$$

Combining this with Eq.(45), and using $M_j = \hbar\omega_j$, gives:

$$\int d^3 r\, \eta(\mathbf{r},\omega_l) R(\mathbf{r},\omega_l) \mathbf{D}_l^*(\mathbf{r}) \cdot \mathbf{D}_j(\mathbf{r}) \approx \frac{1}{2} \hbar\omega_j\, \delta_{lj} \tag{87}$$

This approximate relation can be used to project a mode amplitude from the total field given by Eq.(27), as:

$$\int d^3 r\, \eta(\mathbf{r},\omega_l) R(\mathbf{r},\omega_l) \mathbf{D}_l^*(\mathbf{r}) \cdot \mathbf{D}(\mathbf{r}) =$$

$$\int d^3 r\, \eta(\mathbf{r},\omega_l) R(\mathbf{r},\omega_l) \mathbf{D}_l^*(\mathbf{r}) \cdot \sum_{j=-\infty}^{\infty} \alpha_j e^{-i\omega_j t} \mathbf{D}_j(\mathbf{r}) \approx \frac{1}{2} \hbar\omega_l \alpha_l e^{-i\omega_l t} \tag{88}$$

(Note: We cannot define useful bi-orthogonal 'dual modes' by absorbing $\eta(\mathbf{r},\omega_l) R(\mathbf{r},\omega_l)$ into the mode definition, because such a new mode would not obey the Maxwell equations.)

## 14. The Band Approximation

Returning to structured media, here we show that in a weakly dispersive, weakly structured medium, as is common in wave-guiding media, we can simplify the general theory to a form that is sufficiently accurate and easy to work with. Again, assume the fields propagating in the medium are in distinct spectral bands, denoted as $B_J$, centered at separated center frequencies $\bar{\omega}_J$, with $J = 1,2,3....$ The first step is to make the band approximation in a more general form than done in Eq.(78), so Eqs.(58) and (60) become:

$$\hat{\mathbf{D}}^{J(+)}(\mathbf{r},t) = i\sum_{j\in J}^{\infty} \sqrt{\frac{\varepsilon_0 \hbar\omega_j}{2}} \hat{a}_j e^{-i\omega_j t} \mathbf{u}_j(\mathbf{r})$$

$$\hat{\mathbf{B}}^{J(+)}(\mathbf{r},t) = \frac{i}{c}\sum_{j\in J}^{\infty} \sqrt{\frac{\hbar\omega_j}{2\varepsilon_0}} \hat{a}_j e^{-i\omega_j t} \tilde{\mathbf{u}}_j(\mathbf{r}) \tag{89}$$

with superscript $J$ denoting the band. The second step is to approximate the factors inside Eq.(87) as constant within a given band, so:

$$\frac{1}{\varepsilon_0} \int d^3 r \, \rho_J(\mathbf{r}) \mathbf{D}_l^*(\mathbf{r}) \cdot \mathbf{D}_j(\mathbf{r}) \approx \frac{1}{2} \hbar \omega_j \, \delta_{lj} \tag{90}$$

where we now do have a well-defined weight function for the orthogonality integral in a band:

$$\rho_J(\mathbf{r}) \equiv \varepsilon_0 \eta(\mathbf{r}, \overline{\omega}_J) R(\mathbf{r}, \overline{\omega}_J) \; . \tag{91}$$

In terms of the normalized modes defined in Eq.(56), we have:

$$\int d^3 r \, \rho_J(\mathbf{r}) \mathbf{u}_l^*(\mathbf{r}) \cdot \mathbf{u}_j(\mathbf{r}) \approx \delta_{lj} \tag{92}$$

Then the projection integral Eq.(88) becomes:

$$\frac{1}{\varepsilon_0} \int d^3 r \, \rho_J(\mathbf{r}) \mathbf{D}_l^*(\mathbf{r}) \cdot \hat{\mathbf{D}}^{J(+)}(\mathbf{r}, t) =$$

$$\int d^3 r \, \rho_J(\mathbf{r}) \sqrt{\frac{\hbar \omega_l}{2}} \, \mathbf{u}_l^*(\mathbf{r}) \cdot \sum_{j \in J}^{\infty} \sqrt{\frac{\hbar \omega_j}{2}} \, \hat{a}_j e^{-i\omega_j t} \, \mathbf{u}_j(\mathbf{r}) = \frac{1}{2} \hbar \omega_j \, \hat{a}_j e^{-i\omega_j t} \tag{93}$$

And similarly for $\hat{\mathbf{B}}^{J(+)}(\mathbf{r}, t)$.

We now have a reasonably accurate theory for a weakly dispersive, weakly structured medium, where the modes within a band are orthogonal under a weight function that is well defined in a specific band, making it analogous to the Sturm-Liouville theory discussed earlier. The modes are not precisely orthogonal between frequency-distant bands, because the weight function is different in each band. (We could put a superscript $J$ on the mode functions, but we omit it for simplicity, and try to remember which band we are dealing with.)

## 15. Quantum Fields in Waveguides

Consider an ideal, nonbirefringent dielectric waveguide of length $L$, which tends to infinity, in which the spatial dependence of the weakly dispersive susceptibility depends only on transverse coordinates $\mathbf{x} = (x, y)$, and not on the longitudinal coordinate $z$. We can write the mode functions as $\mathbf{u}_{jm}(\mathbf{r}) = \mathbf{w}_{jm}(\mathbf{x}, \beta_j) L^{-1/2} \exp(i\beta_j z)$, where $\mathbf{w}_{jm}(\mathbf{x}, \beta_j)$ are transverse modes depending only on $(x, y)$. Periodic boundary conditions require the longitudinal propagation constant to have equally spaced values $\beta_j = j 2\pi / L$. For each value of the longitudinal propagation constant $\beta_j$ there exists a discrete set of transverse modes, indexed by integers $m$, which generally have different frequencies denoted $\omega_{jm} = \omega_m(\beta_j)$.

These modes are normalized according to Eq.(92), with the weight function $\rho_J(\mathbf{r})$ replaced by $\rho_J(\mathbf{x}) = \varepsilon_0 \eta_J(\mathbf{x}) R_J(\mathbf{x})$. That is,

$$\int_{-L/2}^{L/2} \frac{dz}{L} \exp[-i(\beta_l - \beta_j)z)] \int d^2 x \, \rho_J(\mathbf{x}) \mathbf{w}_{ln}^*(\mathbf{x}, \beta_l) \cdot \mathbf{w}_{jm}(\mathbf{x}, \beta_j) = \delta_{(l,n)(j,m)} \tag{94}$$

which equals zero unless $(l,n) = (j,m)$. Given that $\beta_j = j2\pi / L$, the $z$ integral equals zero for $j \neq l$. Therefore, two transverse mode functions associated with different values of $\beta_j$ are not necessarily orthogonal in $\mathbf{x}$.

We continue to assume the fields are separated into distinct frequency bands, labeled $B_J$, centered at separated center frequencies $\overline{\omega}_J$, with $J = 1,2,3\dots$. From Eq.(89), the $\mathbf{D}$ field in the $B_J$ band is:

$$\hat{\mathbf{D}}^{J(+)}(\mathbf{r},t) = i \sum_{j \in J}^{\infty} \sum_m \sqrt{\frac{\varepsilon_0 \hbar \omega_j}{2}}\, \hat{a}_{jm} e^{-i\omega_{jm}t}\, \mathbf{w}_{jm}(\mathbf{x},\beta_j) L^{-1/2} \exp(i\beta_j z) \qquad (95)$$

As earlier, in the continuum limit we replace:

$$\sum_j \rightarrow \frac{L}{2\pi} \int_{-\infty}^{\infty} d\beta \quad , \quad \beta_j \rightarrow \beta \ , \ \omega_{jm} \rightarrow \omega_m(\beta)$$

$$\left[\hat{a}_{jm}, \hat{a}_{j'm'}^\dagger\right] = \delta_{jj'} \delta_{mm'} \Rightarrow \frac{L}{2\pi} \int_{-\infty}^{\infty} d\beta \left[\hat{a}(\beta), \hat{a}^\dagger(\beta')\right] \delta_{mm'} = \delta_{mm'}$$

$$\hat{a}_{jm} \doteq \sqrt{L^{-1}}\, \hat{b}_m(\beta)$$

$$\left[\hat{b}_m(\beta), \hat{b}_{m'}^\dagger(\beta')\right] = 2\pi\delta(\beta - \beta')\delta_{mm'}$$

$$(96)$$

Then the displacement field operator in Band $J$ is:

$$\hat{\mathbf{D}}^{J(+)}(\mathbf{r},t)_{Band\,J} = i \sum_m \int_{B_J} \frac{d\beta}{2\pi} \sqrt{\frac{\varepsilon_0 \hbar \omega(\beta)}{2}}\, \hat{b}_m(\beta) e^{-i\omega_m(\beta)t}\, \mathbf{w}_m(\mathbf{x},\beta) \exp(i\beta z) \qquad (97)$$

Note that, as in the plane-wave case earlier, $L$ no longer appears.

Then the orthogonality relation for two modes with possibly different values of $\beta$ becomes, from Eq.(92):

$$\int_{-L/2}^{L/2} \frac{dz}{L} \exp[-i(\beta_l - \beta_j)z)] \int d^2x\, \rho_J(\mathbf{x}) \mathbf{w}_{ln}^{\ *}(\mathbf{x},\beta_l) \cdot \mathbf{w}_{jm}(\mathbf{x},\beta_j) = \delta_{(l,n),(j,m)} \Rightarrow$$

$$\int_{-L/2}^{L/2} \frac{dz}{L} \exp(-i(\beta'-\beta)z) \int d^2x\, \rho_J(\mathbf{x}) \mathbf{w}_n^*(\mathbf{x},\beta') \cdot \mathbf{w}_m(\mathbf{x},\beta) = 2\pi\delta(\beta'-\beta)\delta_{nm}$$

$$(98)$$

Therefore, the transverse modes with equal $\beta$ are orthonormal in two-dimensions:

$$\int d^2x\, \rho_J(\mathbf{x}) \mathbf{w}_n^*(\mathbf{x},\beta) \cdot \mathbf{w}_m(\mathbf{x},\beta) = \delta_{nm} \qquad (99)$$

Note the weight function $\rho_J(\mathbf{x})$ contains information about the material dispersion. From Eq.(46):

$$\rho_J(\mathbf{x}) = \varepsilon_0 \eta_J(\mathbf{x}) R_J(\mathbf{x})$$

$$= \varepsilon_0 \eta_J(\mathbf{x}) \frac{\mathrm{v}_p(\mathbf{r}, \overline{\omega}_J)}{\mathrm{v}_g(\mathbf{r}, \overline{\omega}_J)} \tag{100}$$

$$= \varepsilon_0 \left\{ \eta(\mathbf{r}, \omega) - \frac{\omega}{2} \frac{\partial \eta(\mathbf{r}, \omega)}{\partial \omega} \right\}_{\omega = \overline{\omega}_J}$$

Eq.(46) also includes the form:

$$R(\mathbf{r}, \omega) = \frac{\mathrm{v}_p(\mathbf{r}, \omega)}{\mathrm{v}_g(\mathbf{r}, \omega)} = \frac{1}{n(\mathbf{r}, \omega)} \frac{\partial}{\partial \omega} \big( \omega \, n(\mathbf{r}, \omega) \big) \tag{101}$$

which shows that in many typical media, wherein refractive index is proportional to material density, the spatial dependence of this density cancels, and $R(\mathbf{r}, \omega)$ is approximately independent of position $\mathbf{r}$, so $R(\mathbf{r}, \omega) \approx R(\omega) \approx R_J$. In such cases, we can approximate:

$$\int d^2 x \, \varepsilon_0 \eta_J(\mathbf{x}) \mathbf{w}^*_{\,n}(\mathbf{x}, \beta) \cdot \mathbf{w}_m(\mathbf{x}, \beta) = \frac{1}{R_J} \delta_{nm} = \frac{\mathrm{v}_g(\overline{\omega}_J)}{\mathrm{v}_p(\overline{\omega}_J)} \delta_{nm} \tag{102}$$

The ratio of velocities could be absorbed into the modes definition and consequently the factors appearing inside the integral in Eq.(97). But remember that the phase and group velocities here refer only to those properties that would exist in the bulk material, and they do not include effects of waveguiding, which we discuss next.

## 16. Waveguide Modes

To solve for the modes of an arbitrary waveguide formed by a structured dielectric, it is convenient to return to the Maxwell equations, Eqs.(29),(30). See (1, 28)

$$i\omega_J \mathbf{B}_J(\mathbf{r}) = \nabla \times \left[ \eta(\mathbf{r}, \omega_J) \mathbf{D}_J(\mathbf{r}) \right] \tag{103}$$

$$-i\omega_J \mu_0 \mathbf{D}_J(\mathbf{r}) = \nabla \times \mathbf{B}_J(\mathbf{r}) \tag{104}$$

Then use vector identities:

$$\nabla \times \eta \mathbf{A} = \eta \nabla \times \mathbf{A} + \nabla \eta \times \mathbf{A},$$

$$\nabla \times \nabla \times \mathbf{A} = \nabla \big( \nabla \cdot \mathbf{A} \big) - \nabla^2 \mathbf{A} \tag{105}$$

And the Maxwell equation, $\nabla \cdot \mathbf{B}_J = 0$, to find:

$$\nabla \times \left[ \eta(\mathbf{r}, \omega_j) \nabla \times \mathbf{B}_j(\mathbf{r}) \right] = \mu_0 \omega_j^{\,2} \mathbf{B}_j(\mathbf{r})$$

$$\eta \nabla \times \nabla \times \mathbf{B}_j + \nabla \eta \times \left( \nabla \times \mathbf{B}_j \right) = \mu_0 \omega_j^{\,2} \mathbf{B}_j(\mathbf{r})$$

$$-\eta \nabla^2 \mathbf{B}_j + \nabla \eta \times \left( \nabla \times \mathbf{B}_j \right) = \mu_0 \omega_j^{\,2} \mathbf{B}_j(\mathbf{r}) \tag{106}$$

$$\nabla^2 \mathbf{B}_j + \frac{\mu_0 \omega_j^{\,2}}{\eta} \mathbf{B}_j(\mathbf{r}) = \frac{\nabla \eta}{\eta} \times \left( \nabla \times \mathbf{B}_j \right)$$

$$\nabla^2 \mathbf{B}_j + \varepsilon(\mathbf{r}, \omega_j) \mu_0 \omega_j^{\,2} \mathbf{B}_j = -\nabla \ln[\varepsilon(\mathbf{r}, \omega_j)] \times \left( \nabla \times \mathbf{B}_j \right)$$

Here we have restored the frequency dependence of the permittivity, in order to account for material dispersion within a band. Thus, we are going slightly beyond the approximation used in the earlier treatment of orthogonality. To be consistent, we have to replace $\varepsilon(\mathbf{r}, \omega_j) \rightarrow \varepsilon(\mathbf{r}, \overline{\omega}_j)$.

Once the $B$ modes are determined by solving Eq.(106), the $D$ modes are given by:

$$\mathbf{D}_j(\mathbf{r}) = \frac{i}{\omega_j \mu_0} \nabla \times \mathbf{B}_j(\mathbf{r}) \tag{107}$$

In many cases, the right-hand side of Eq.(106) is negligible to good approximation. In a piece-wise medium, made of discrete regions with uniform $\varepsilon(\mathbf{r}, \omega_j)$ connected at sharp interfaces, the right-hand side can be set to zero, and its effects are manifested in terms of boundary conditions at the interfaces. In continuous-index media, where the right-hand side is not strictly zero, it creates only small perturbations to mode propagation, such as 'spin-orbit coupling' (29, 30). So for simplicity, here we approximate it as zero.

Now, consider an ideal dielectric waveguide of length $L$, in which the spatial dependence of the weakly dispersive susceptibility $\varepsilon(\mathbf{x}, \omega_j)$ is transverse only, as in the previous section. Also make the band approximation, in which the medium has frequency-independent $\varepsilon_j(\mathbf{x})$ throughout the chosen band. Because the medium depends only on transverse coordinates, we can write the scaled magnetic-field modes in separable form:

$$\tilde{\mathbf{u}}_{jm}(\mathbf{r}) = \tilde{\mathbf{w}}_{jm}(\mathbf{x}) L^{-1/2} \exp(i\beta_j z) \tag{108}$$

where the $m$ index labels the transverse function. Then, since the scaled modes $\tilde{\mathbf{u}}_{jm}(\mathbf{r})$ are proportional to the $\mathbf{B}_j(\mathbf{r})$ modes, they also obey:

$$\nabla^2 \tilde{\mathbf{u}}_{jm} + \varepsilon(\mathbf{x}, \omega_{jm}) \mu_0 \omega_{jm}^2 \tilde{\mathbf{u}}_{jm} = RHS \tag{109}$$

where $RHS \doteq -\nabla \ln[\varepsilon(\mathbf{x}, \omega_{jm})] \times (\nabla \times \tilde{\mathbf{u}}_{jm}) \approx 0$. Thus, noting that the square of the material refractive index is $n^2(\mathbf{x}, \omega_{jm}) = \varepsilon(\mathbf{x}, \omega_{jm}) / \varepsilon_0$, gives:

$$\nabla_{\mathbf{x}}^{\,2} \mathbf{w}_{jm}(\mathbf{x}, \beta_j) + \left( \omega_{jm}^2 c^{-2} n^2(\mathbf{x}, \omega_{jm}) - \beta_j^{\,2} \right) \mathbf{w}_{jm}(\mathbf{x}, \beta_j) \approx 0 \tag{110}$$

Again, periodic boundary conditions in $z$ require the propagation constant to have equally spaced values $\beta_j = j2\pi / L$. For each value $\beta_j$ there exists a set of transverse modes, indexed by integer $m$, with frequencies $\omega_{jm} = \omega_m(\beta_j)$, determined by solving Eq.(110). Then in the continuum limit, $\beta_j \to \beta$, $\omega_{jm} \to \omega_m(\beta)$, we have:

$$\nabla_{\mathbf{x}}^2 \mathbf{w}_m(\mathbf{x}, \beta) + \left( \omega_m^2(\beta) c^{-2} n^2(\mathbf{x}, \omega_m) - \beta^2 \right) \mathbf{w}_m(\mathbf{x}, \beta) = 0 \qquad (111)$$

Thus, for every (continuous) value of $\beta$, there exists a set of transverse modes $\mathbf{w}_m(\mathbf{x}, \beta)$.

Now, being in the continuum limit, we may invert the relation between frequency and propagation constant, and consider frequency as the independent variable ($\omega_m \to \omega$) such that $\beta_m = \beta_m(\omega)$. And rewrite the wave equation as an eigenvalue problem:

$$\nabla_{\mathbf{x}}^2 \mathbf{w}_m(\mathbf{x}) + \omega^2 c^{-2} n^2(\mathbf{x}, \omega) \mathbf{w}_m(\mathbf{x}) = \beta_m^2 \mathbf{w}_m(\mathbf{x}) \qquad (\omega \ fixed) \qquad (112)$$

This equation has the form of the 2D Schroedinger equation. Defining $V(\mathbf{x}) \equiv -\omega^2 c^{-2} n^2(\mathbf{x}, \omega)$, $E_m \equiv -\beta_m^2$, we have:

$$-\nabla_{\mathbf{x}}^2 \mathbf{w}_m(\mathbf{x}) + V(\mathbf{x}) \mathbf{w}_m(\mathbf{x}) = E_m \mathbf{w}_m(\mathbf{x}) \qquad (113)$$

This Schroedinger equation analogy can give us insights into the nature of the bound states of light in a waveguide.

## 17. Example of Waveguide Modes

Consider an idealized case for illustration: an infinite-barrier, 2D slab waveguide. The susceptibility varies only in the $x$ direction, and we write the wave equation Eq.(112) for only one transverse vector component, denoted as $u_m(x)$:

$$\partial_x^2 u_m(x) + \omega^2 c^{-2} n^2(x, \omega) u_m(x) = \beta_m^2 u_m(x) \qquad (114)$$

The boundary conditions are $u_m(x = \pm D / 2) = 0$, where $D$ is the slab thickness. Solutions are:

$$u_m(x) = A\cos(k_x x) \ or \ A\sin(k_x x) \qquad (115)$$

with $k_x = m\pi / D$, and $-k_x^2 + \omega^2 n^2(\omega) / c^2 = \beta_m^2$, thus:

$$\beta_m = \sqrt{\omega^2 n^2(\omega) / c^2 - (m\pi / D)^2} \qquad (116)$$

Figure 1 shows an example plot of $\beta_m$ vs $\omega$, for $m = 0,1,2,3,\ldots$

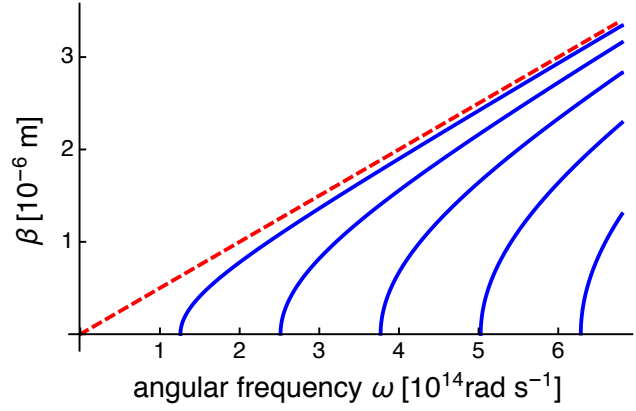

Figure 1. Propagation constants $\beta_m$ versus frequency $\omega$ for different transverse modes, indexed by $m$ (from left to right, $m$ = 0,1,2,3,4) The dashed line corresponds to that of the bulk material ($j$=0). Parameters used: $n(\omega) = 1.5$, $D = 5\,\mu m$.

According to the solutions given, higher-order modes have slower group velocity than lower-order ones, consistent with the light traveling in steeper zigzag paths. This result can be seen by the following calculations, in which we denote $k \doteq k(\omega) \doteq \omega\, n(\omega) / c$ so that

$k(\omega)^2 = \omega^2 n^2(\omega) / c^2 = \beta^2 + k_x^{\,2}$. Define $\kappa$ and $\cos(\theta)$ by:

$$\beta = \sqrt{k(\omega)^2 - (m\pi / D)^2} \doteq \kappa \cos(\theta) \qquad (117)$$

where

$$\cos(\theta) = \frac{\sqrt{k(\omega)^2 - (m\pi / D)^2}}{k(\omega)} = \sqrt{1 - k(\omega)^{-2}(m\pi / D)^2} \ \ \leq 1 \qquad (118)$$

So, $\theta$ plays the role of the angle defining a zigzag path. Then $\beta^2 = k(\omega)^2 - (m\pi / D)^2$ implies that, if $n$ is frequency-independent within a band, we have:

$$\frac{d\beta^2}{d\omega} = 2\frac{\omega n^2}{c^2} = 2\beta\frac{d\beta}{d\omega} \qquad (119)$$

And so the group velocity is given by

$$\frac{1}{v_g} = \frac{d\beta}{d\omega} = \frac{1}{\beta}\frac{\omega n^2}{c^2} = \frac{n}{c}\frac{k}{\sqrt{k^2 - (m\pi / D)^2}}$$

$$v_g = \frac{c}{nk}\sqrt{k^2 - (m\pi / D)^2} = \frac{c}{n}\cos(\theta) \ < c \qquad (120)$$

and the phase velocity by

$$\mathrm{v}_p = \frac{\omega}{\beta} = \frac{\omega}{\sqrt{k^2 - (m\pi / D)^2}} = \frac{c}{n} \frac{1}{\sqrt{1 - (2\pi)^{-2}(m\pi \lambda / D)^2}} > \frac{c}{n}$$

$$\mathrm{v}_p = \frac{c}{n} \frac{1}{\cos(\theta)} > \frac{c}{n}$$

(121)

If material dispersion is included, then the resulting modifications to the phase and group velocities can be calculated from the frequency-dependent solutions $\beta_m(\omega)$ of Eq.(112). It is found typically that for waveguides with width greater than several wavelengths the material dispersion dominates, with velocities given by Eq.(14):

$$\frac{1}{\mathrm{v}_g} = \frac{\partial k}{\partial \omega} = \frac{1}{c}\left( n(\omega) + \omega \frac{\partial n}{\partial \omega} \right)$$

$$\mathrm{v}_p = \frac{c}{n(\omega)}$$

(122)

while for smaller widths the waveguide influence begins to dominate, as in Eqs.(120) and (121).

Waveguides may also have structure in the longitudinal, $z$, direction, forming Bragg gratings for dispersion engineering, or resonant cavities for cavity modification of quantum optics processes such as spontaneous parametric down conversion (16). The methods given here can easily be adopted to such structured waveguides.

## 18. Concluding Remarks

These self-contained and tutorial notes explored a direct, macroscopic approach to quantizing a linear-response dielectric material with both spectral dispersion and spatial nonuniformity, and the spectral region of interest is optically transparent so that explicit treatment of the underlying dynamical physics of the medium is not needed. The approach taken is based on the macroscopic Maxwell equations and a corresponding Hamiltonian, without Lagrangians, and uses a standard mode-based quantization method. New results include a direct derivation of the mode normalization condition, and a direct proof of the nonorthogonality of the modes.

The main results derived are:

a. The energy density and flux in a dispersive dielectric. (Sec. 3)
b. An illustration of energy density and flux for a plane wave of incoherent, stationary light traveling in the z-direction in a *nonstructured* medium. (Sec. 3)
c. A summary of the inverse permittivity formalism. (Sec. 4)
d. A new derivation of the mode normalization and nonorthogonality conditions. (Sec. 5)
e. A straightforward quantization scheme using only the above-mentioned results and the Hamiltonian for the system. (Sec. 6)
f. Examples of the quantized field expressions for various special cases. (Secs. 7, 8, and 10)
g. Verification of the relationship between group velocity and energy flux. (Sec. 9)
h. The band approximation and the continuum limit. (Secs. 11 and 14)
i. Approximate orthogonality relations. (Sec.13)
j. Treatment of waveguide modes. (Secs. 15, 16 and 17)

The author knows of no other published derivations of all of these results from a purely macroscopic Hamiltonian formalism without the use of dynamical models for the medium.

**Disclosure statement.** The author has no financial interest or benefit that has arisen from the direct applications of this research.

## Acknowledgements

The author thanks Peter Drummond, Peter Milonni, and John Sipe for many very helpful discussions, and the National Science Foundation for support under grants PHY-1820789 and PHY-1839216.